\documentstyle[11pt,aaspp4]{article} 
\input psfig
\begin{document}
\title{Strong Relativistic Collisionless Shock} 
\author{Andrei Gruzinov}
\affil{Physics Department, New York University, 4 Washington Place, New York, NY 10003}

\begin{abstract}
A direct numerical simulation of a strong relativistic collisionless shock propagating into an unmagnetized medium has been performed in two spatial dimensions. It is found that: (i) collisionless shock exists, (ii) particle acceleration is insignificant, (iii) the shock does not generate ``quasi-static'' magnetic fields, contrary to recent claims by other authors. All our results agree with a simple plasma-physical model of collisionless shocks. On the other hand, astrophysical evidence indicates that shocks do accelerate charged particles, and GRB phenomenology indicates that shocks do generate strong magnetic fields. The conflict between the simplest plasma-physical model and astronomical observations of collisionless shocks is discussed. 

\end{abstract}
\keywords{magnetic fields -- shock waves -- acceleration of particles}

\section{Strong relativistic shocks}

We will discuss strong relativistic shock waves. These shocks should generate magnetic fields and accelerate charged particles in radio supernovas (Chevalier 1982,  Fransson \& Bjornsson 1998) and gamma-ray bursts (Sari, Piran, Narayan 1997, Gruzinov \& Waxman 1998, Gruzinov 2001). 

The magnetic field of the unshocked plasma is probably insignificant in radio supernovas, and is certainly negligible in GRB afterglows. For GRB afterglows, for example, magnetic energy in the shocked plasma is of order GeV per particle. Magnetic energy must be less than $T$ per particle in the unshocked plasma, where $T$ is the temperature of the unshocked plasma. Therefore simple shock compression of the pre-existing field is irrelevant, the magnetic field should be generated by the shock (see Gruzinov 2001 for a detailed discussion). For all practical purposes, the unshocked plasma is cold and unmagnetized.

The shocks are collisionless (Sagdeev 1966 gives a good introduction), meaning that the shock transition is accomplished by electromagnetic fields, rather than particle collisions. These shocks are turbulent, involving fluctuating electromagnetic fields and particle distribution functions. There are only two reliable tools for studying such shocks: numerical simulations and common sense. 

\begin{figure}[htb]
\psfig{figure=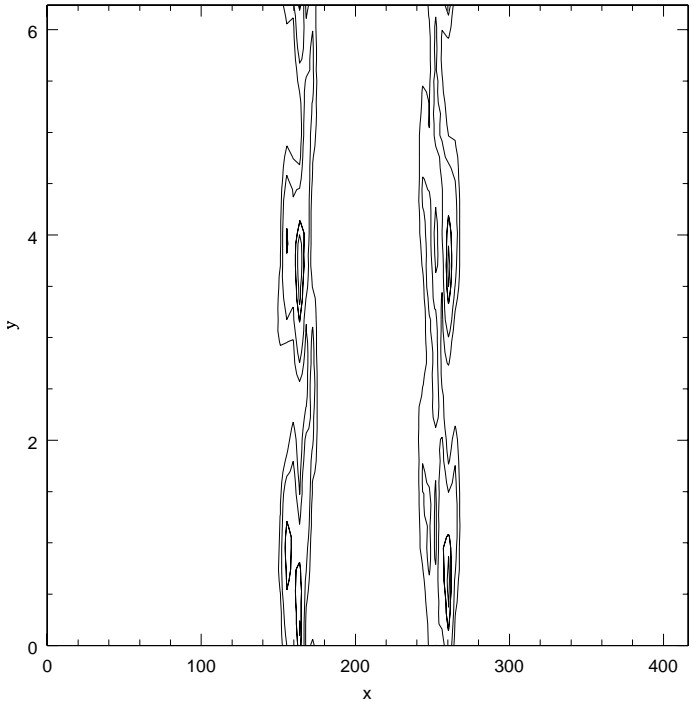,width=4in}
\caption{Magnetic field contours at $\omega _pt=208$. Box sizes are in units of skin depth. At t=0, unmagnetized plasma in the left half of the box moved to the right at $v/c=0.56$, the right half moved to the left at the same speed. It is seen that magnetic field is concentrated near the two shock fronts. The plasma in the center of the box is demagnetized.}
\end{figure}

A direct numerical simulation in two spatial dimensions (Kazimura et al 1998) has apparently demonstrated: ``generation of a small-scale quasi-static magnetic field''. And also that ``particles can be accelerated ``. We have repeated the simulation, and have reproduced the results. But our run has been longer, and our conclusions are different: (ii) no significant particle acceleration is seen, (iii) the shock does not generate ``quasi-static'' magnetic fields. Another important finding, which agrees with Kazimura et al (1998) simulation is that (i) collisionless shock exists \footnote{This is not an empty statement. The very concept of self-excited electromagnetic field mediating a shock transition needs to be confirmed for a relativistic shock propagating into unmagnetized plasma.}.

In \S 2 we define the problem, in \S 3 we present our numerical results, in \S 4 we discuss the conflict between plasma-physical and astronomical understanding of strong relativistic collisionless shocks.

\section{Plasma Physics and Astrophysics of Strong Relativistic Collisionless Shocks}

The following qualitative description of a collisionless relativistic shock propagating into a cold unmagnetized plasma seems probable from a plasma-physical point of few. Neglect, just to simplify the argument, the difference between electron and ion masses, or think of collisionless electron-positron plasma. Also, assume that the shock is mildly relativistic. Than there are only three dimensional parameters in the problem (plasma mass density $nm$, plasma charge density $ne$, and the speed of light $c$), and therefore there are no dimensionless parameters \footnote{In collisionless plasma, described by Vlasov equations, that is Maxwell for electromagnetic field plus Boltzmann for the distribution function, the plasma parameters $e$, $m$, $n$ enter only in combinations $ne$ and $nm$.}. 

The shock transition region should then be just few $\delta$ thick, where $\delta = c/\omega _p$ is the only available length scale -- collisionless skin depth; $\omega _p=(4\pi ne^2/m)^{1/2}$ is the plasma frequency. Within the transition region, anisotropy of the distribution function excites kinetic instabilities (Weibel, two-stream, see Sagdeev 1966). These instabilities generate strong (equipartition, $B^2\sim nmc^2$), small-scale ($\sim \delta$) electromagnetic fields. The fields mediate the shock transition: they isotropize the distribution function, and bring the plasma up the shock adiabatic. After the transition is accomplished, that is a few $\delta$ downstream, the fields decay by Landau damping. Our numerical simulations, \S 3, fully confirm this simple picture (in 2 dimensions).

Now, assuming that the above description of a strong relativistic collisionless shock is correct, let us see what are the astrophysical consequences. Since the fields exist only in a few-skin-thick layer, high energy particles cannot be confined, and therefore there is no particle acceleration. For the same reason, synchrotron emission from such a shock is insignificant.

The accepted astrophysical scenario is very different: The shock generates near equipartition magnetic fields and accelerates charged particles. This leads to synchrotron emission characterized by a power law at high frequencies. Both radio-supernovas (Fransson \& Bjornsson 1998) and gamma-ray bursts (Sari, Piran, Narayan 1997) can be explained under this assumption. The generated magnetic field can survive only if it somehow propagates to large scales (Gruzinov \& Waxman 1998, Gruzinov 2001). We do not have a clear picture of how this might happen.

\section{Numerical simulation using TRISTAN}

We performed direct numerical simulations of relativistic collisionless shocks in two spatial dimensions. The geometry of the simulation can be described as follows. The particles move in the $(x,y)$ plane. The electric field is $(E_x(x,y,t),E_y(x,y,t),0)$, the magnetic field is $(0,0,B_z(x,y,t))$. 

The simulation was performed using a 2D version of the code TRISTAN (Buneman 1993) (TRISTAN in what follows). This code was also used by Kazimura et al (1998) (K98 in what follows). A large number of simulations was performed, to study numerical accuracy and the dependence on physical parameters. In all simulations, the initial electromagnetic field was zero. The initial plasma velocities were mildly relativistic in all our runs. We simulated both hot and initially cold plasma, and obtained similar results. 

The run presented in Figures 1 and 2 (thick lines in Fig. 2) uses approximately the same TRISTAN parameters as K98. We choose to present this particular run to compare the results. The simulation parameters (in K98 notations, see also TRISTAN) are as follows. The system size is $L_y=60$, $L_x=4000$, the grid size is 1. K98 have $L_x=1024$, and correspondingly can only afford a shorter run time. This is the main difference between the two simulations. Periodic boundary conditions on particles and fields are used. Particle number per cell is 16 (1 for thin lines in Fig. 2). Initial plasma velocity is  $v_x=0.56c$ at $x<L_x/2$, and $v_x=-0.56c$ at $x>L_x/2$. The speed of light $c=0.5$ (see TRISTAN). Thermal velocity $v_T=0.1c$. Plasma frequency $\omega _p=.052$, positive and negative particles have equal mass.

\begin{figure}[htb]
\psfig{figure=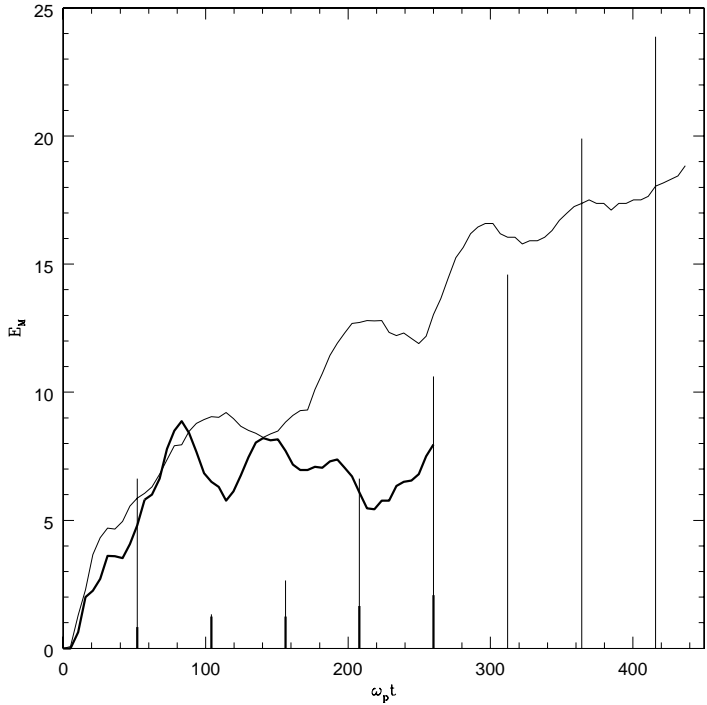,width=4in}
\caption{The generation of magnetic energy. Thin lines -- poor accuracy. Thick lines -- better accuracy. Horizontal axis is the time multiplied by the plasma frequency. Vertical axis is the magnetic energy in units of $6.24nmv^2\delta ^2$, where $n$ is the density, $m$ is the mass, $v$ is the collision speed of the two parts of the plasma, $\delta$ is the skin depth, $6.24\delta$ is the width of the simulation box, see Fig.1. Vertical lines correspond to numerical error; the lines give absolute value of the total energy change as a function of time (in units of $6.24nmv^2\delta ^2$). }
\end{figure}

Our numerical simulation fully supports the simple plasma-physical picture of \S 2. Magnetic fields exist only around the shock fronts, Fig. 1. Total magnetic energy saturates. The distribution function does not develop a power-law tail, that is particle acceleration is insignificant (although a small number of particles is found to have Lorentz factors $\sim 10$).

Also, our simulation results are in agreement with the results (but not the conclusions !) of K98. At small times, about as long as K98 ran their simulation, magnetic energy grows at an approximately constant rate, giving an impression that there is a stationary process of converting kinetic into magnetic energy. Indeed, both bad and good accuracy curves in Fig.2 show a kinetic to magnetic energy conversion of order few percent at time $\omega _pt<80$, a similar conversion rate is reported by K98. However, at later time, the saturation is clearly seen in Fig.2. Fig. 1 explains why: magnetic energy exists only in the vicinity of the shock fronts. Plasma is magnetized by the shock passage, but later on it demagnetizes.

It might seem surprising, that such a long numerical run is needed to see the saturation effect. One expects that a few plasma times should be enough. Reality (numerical)  is a little more difficult. Since even near the shock fronts magnetic energy is noticeably below equipartition (few percent), the Larmor radius is a few times larger than the skin depth. It takes a few Larmor orbits to damp the field, so the shock front gets inflated to some tens of skin depths. But the important thing is that in the long run the field does decay. 

\section{Discussion}

We have seen that a relativistic collisionless shock propagating into an unmagnetized medium does not generate magnetic fields and does not accelerate charged particles. This is in contradiction with the GRB phenomenology , assuming that afterglows are synchrotron (Gruzinov 2001). 

This author believes, without any sound reason (other than GRB afterglow phenomenology), that the astrophysical rather than our simple plasma-physical picture is indeed correct: Shocks generate magnetic fields and accelerate charged particles. 

  If so, what's missing from our numerical simulation? First, propagation of the fields to large scales is a form of dynamo, and dynamo requires three spatial dimensions. But this might not be enough. Our preliminary 3D simulations show magnetic field evolution very similar to 2D. It might be that external large-scale inhomogeneities (non-uniform shock, non-uniform interstellar medium, etc.) are required. This external inhomogeneity requirement makes the model ugly, but it might be the correct one.

Direct numerical simulations of 3D collisionless relativistic shocks is a doable, and a very interesting problem. If performed, it would yield, for the first time, a real understanding of particle acceleration by relativistic shocks (Monte Carlos with some assumed scatterers is an untrustworthy toy model).

\end{document}